\documentclass[a4j,11pt,onecolumn,oneside,notitlepage,final]{article}

\usepackage{graphicx}
\usepackage{xcolor}
\usepackage[caption=false]{subfig}
\usepackage{mathrsfs,mathtools}
\usepackage{
amssymb}
\usepackage{physics}
\usepackage{bm}
\usepackage{braket}
\usepackage{listings}
\usepackage{cases}
\usepackage{comment}
\usepackage{soul}
\usepackage{cancel}
\usepackage{cases}
\usepackage[utf8]{inputenc}
\usepackage{url}
\usepackage{longtable}
\usepackage[normalem]{ulem}
\usepackage{xspace}
\usepackage{slashed}
\usepackage[colorlinks=true
,urlcolor=darkblue
,anchorcolor=darkblue
,citecolor=darkblue
,filecolor=darkblue
,linkcolor=darkblue
,menucolor=darkblue
,linktocpage=true
,pdfproducer=medialab
,pdfa=true
]{hyperref}
\usepackage{cite}

\newcommand{\ns}{n_\mathrm{s}}

\newcommand{\fNL}{f_\mathrm{NL}}
\newcommand{\tNL}{\tau_\mathrm{NL}}
\newcommand{\bfNL}{\bar{f}_\mathrm{NL}}
\newcommand{\btNL}{\bar{\tau}_\mathrm{NL}}

\newcommand{\uB}{\mathrm{B}}

\newcommand{\ue}{\mathrm{e}}

\newcommand{\bfk}{\mathbf{k}}

\newcommand{\uL}{\mathrm{L}}

\newcommand{\bfq}{\mathbf{q}}

\newcommand{\uS}{\mathrm{S}}

\newcommand{\bfx}{\mathbf{x}}

\newcommand{\bae}[1]{\begin{align} #1 \end{align}}
\newcommand{\bce}[1]{\begin{cases} #1 \end{cases}}
\newcommand{\dps}{\displaystyle}

\definecolor{monza}{HTML}{CF000F}
\definecolor{darkblue}{HTML}{00008b}
\definecolor{darkmagenta}{HTML}{8b008b}

\def\be{\begin{equation}}
\def\ee{\end{equation}}
\def\ben{\begin{eqnarray}}
\def\een{\end{eqnarray}}

\begin{document}
	\begin{center}
		\vskip .5in
		
		{\Large \bf
			Local observer effect on the cosmological soft theorem
		}
		\vskip .45in
{
Teruaki Suyama$^{a}$,
Yuichiro Tada$^{b}$,
Masahide Yamaguchi$^{a}$
}

{\em
$^a$
   Department of Physics, Tokyo Institute of Technology, 2-12-1 Ookayama, Meguro-ku, Tokyo 152-8551, Japan
}\\
{\em
$^b$
    Department of Physics, Nagoya University, Nagoya 464-8602, Japan
}\\

\end{center}
	
\abstract{
Non-Gaussianities of primordial perturbations in the soft limit provide 
important information about the light degrees of freedom during inflation.
The soft modes of the curvature perturbations, unobservable for a local observer, act to rescale the spatial coordinates.
We determine how the trispectrum in the collapsed limit is shifted by the rescaling due to the soft modes.
We find that the form of the inequality between the $\fNL$ and $\tNL$ parameters is not affected by the rescaling, demonstrating that the role of the inequality as an indicator of the light degrees of freedom remains intact.
We also comment on the local observer effect on the consistency relation for ultra slow-roll inflation.
}

\section{Introduction}

The non-Gaussianity of primordial curvature perturbations has been investigated intensively both theoretically and observationally over the decades as a powerful probe of the physics of inflation
and the generation mechanism of primordial perturbations (e.g. Ref.~\cite{Meerburg:2019qqi}). 
In particular, the non-Gaussian properties of primordial perturbations in the soft limit provide 
important information about the light degrees of freedom (DoF) during inflation.
The three-point function (bispectrum) of the curvature perturbation $\zeta$ in the squeezed limit is commonly represented by the
$\fNL$ parameter as
\be
\fNL=\frac{5}{12}\lim_{k_3\to0}\frac{B_\zeta(k_1,k_2,k_3)}{P_\zeta(k_1)P_\zeta(k_3)}, \label{def-fnl}
\ee
where $P_\zeta$ and $B_\zeta$ are defined by
\bae{
    \bce{
        \dps
        \braket{\zeta_{\bfk}\zeta_{\bfk^\prime}}=(2\pi)^3\delta^{(3)}(\bfk+\bfk^\prime)P_\zeta(k), \\
        \dps
        \braket{\zeta_{\bfk_1}\zeta_{\bfk_2}\zeta_{\bfk_3}}=(2\pi)^3\delta^{(3)}(\bfk_1+\bfk_2+\bfk_3)B_\zeta(k_1,k_2,k_3).
    }
}
In any slow-roll single field inflation model, this parameter is solely written in terms of the
spectral index and the power spectrum as~\cite{Maldacena:2002vr}
\be
    \fNL=\frac{5}{12} (1-\ns), \qq{with $\ns-4=\dv{\log P_\zeta}{\log k}$.} \label{fnl-single}
\ee
If fields other than the inflaton also contribute to the curvature perturbations,
this relation no longer holds and $\fNL$ takes a different value that depends on the details of the underlying model. 
Another example, which is the main target of this paper, is the relation between the collapsed four-point function
and the squeezed three-point function given by~\cite{Suyama:2007bg,Suyama:2010uj} 
\bae{
    \tNL\geq\left(\frac{6}{5}\fNL\right)^2,  \label{f-t-relation}
}
where
\bae{
 \tNL=\frac{1}{4}\lim_{k_{12}\to0}\frac{T_\zeta(\bfk_1,\bfk_2, \bfk_3, \bfk_4)}{P_\zeta(k_1)P_\zeta(k_3)P_\zeta(k_{12})},
 \label{def-tnl}
}
$k_{12}=|\bfk_1+\bfk_2|$, and the trispectrum $T_\zeta$ represents the connected part of the
four-point function
\bae{
        \dps
       \braket{\zeta_{\bfk_1}\zeta_{\bfk_2}\zeta_{\bfk_3}\zeta_{\bfk_4}}=(2\pi)^3\delta^{(3)}(\bfk_1+\bfk_2+\bfk_3+\bfk_4)T_\zeta(\bfk_1,\bfk_2, \bfk_3, \bfk_4)+{\rm disconnected}.
       \label{4-cone}
}
Physically, this $\fNL$-$\tNL$ relation arises from the fact that the correlation between the amplitude of the large-scale curvature perturbations and the variance of the small-scale curvature perturbations, which is represented by $\fNL$, 
inevitably produces large-scale modulation of the variance of the small-scale curvature perturbations, which is represented by $\tNL$, while the opposite is not true in general.
One notable feature of this relation is that it reaches equality when only one DoF contributes to the curvature perturbations,
and $\tNL$ becomes larger than the lower bound otherwise \cite{Suyama:2007bg}. 
Thus, observational confirmation of the violation of the equality 
clearly shows the existence of the extra DoF compared to the adiabatic mode.

On the other hand, the observational aspect of non-Gaussianity also needs careful study because of the non-linearity of gravity itself.
To predict non-Gaussian observables, a time-evolution scheme at the same perturbation order is required in general.
However, if the soft mode 
is a superhorizon, it can be renormalized into the background, thanks to the separated universe assumption, and therefore the required perturbation order decreases by one.
The effective non-linearity (NL) parameter $\bfNL$ on the renormalized background is modified from the original one as~\cite{Pajer:2013ana}
\bae{
    \bfNL=\fNL+\frac{5}{12} (\ns-1),  \label{fnl-cons}
}
which represents the so-called \emph{local observer effect} (LOE)~\cite{Tada:2016pmk}.
In particular, for slow-roll single-field inflation,
inserting Eq.~(\ref{fnl-single}) yields $\bfNL =0$,
demonstrating that the non-vanishing correlation between the soft modes and the hard modes expressed by Eq.~(\ref{fnl-single})
is merely apparent~\cite{Tanaka:2011aj, Pajer:2013ana}.
Given this, a natural question is whether the $\fNL$-$\tNL$ relation (\ref{f-t-relation}) still holds for these effective NL parameters 
\bae{
    \btNL\overset{?}{\geq}\left(\frac{6}{5}\bfNL\right)^2,
}
or not.
Like the relation (\ref{fnl-single}), when the higher-order correlation function in the soft limit is expressed in terms of the tilt of the lower-order correlation functions (see, e.g., \cite{Huang:2006eha}), this will merely reflect the apparent correlation produced by the soft modes, and the form of such a relation will be changed in the rescaled coordinates.
On the other hand, based on the explanation given below Eq.~(\ref{4-cone}),
one may expect that the $\fNL$-$\tNL$ relation in the rescaled coordinates should also hold true.
To the best of our knowledge, no answer to this question has been given in the literature, and
one goal of this short paper is to address the question by
deriving the transformation rule of the trispectrum under the renormalization of the soft modes.

\section{Renormalized bispectrum and trispectrum}

Throughout this paper, we consider a situation in which the curvature perturbation $\zeta$ is generated during/after inflation and remains conserved afterwards when the modes are on super-Hubble scales.
Locally, the soft (long-wavelength) modes $\zeta_\uL (\bfx)$ of the curvature perturbation evaluated
at a point~$\bfx$
\bae{
    \zeta_\uL (\bfx)=\int \frac{\dd[3]{q}}{(2\pi)^3}
    \ue^{i \bfq \cdot \bfx} W_q \zeta_\bfq,
}
where $W_q$ is a window function that extracts only the soft modes,
can be absorbed into the background space,
which amounts to a local rescaling of the spatial coordinates, $\bfx\to\bar{\bfx}=(1+\zeta_\uL(0))\bfx$.
By this rescaling, the curvature perturbation in the vicinity of the origin transforms as (see, e.g., Ref.~\cite{Hinterbichler:2012nm})
\be
    \zeta (\bfx) \to {\bar \zeta}(\bfx)=\zeta (\bfx) 
    - \zeta_\uL (0)(1+ \bfx \cdot 
    \grad{\zeta (\bfx)} ),
\ee
or, in terms of the Fourier modes,
\be
    \zeta_\bfk \to {\bar \zeta}_\bfk=\zeta_\bfk+\zeta_\uL (0) \left( 3+\bfk \cdot \partial_\bfk \right) \zeta_\bfk. \label{zeta-k-transform}
\ee
Using Eq.~(\ref{zeta-k-transform}), the power spectrum of the local curvature perturbation in the presence of the fixed soft modes
can be written as (see Refs.~\cite{Huang:2006eha, Cheung:2007sv, Seery:2008ax})
\be
\langle {\bar \zeta}_{\bfk_1} {\bar \zeta}_{\bfk_2} \rangle_\uB
= \langle \zeta_{\bfk_1} \zeta_{\bfk_2} \rangle_\uB
+(\ns-1) 
\int \frac{\dd[3]{q}}{(2\pi)^3} W_q \zeta_\bfq \,(2\pi)^3 \delta^{(3)} 
(\bfk_1+\bfk_2+\bfq ) P_\zeta (k_1), \label{power-b}
\ee
where $\braket{\cdots}_\uB$ is the local ensemble average on the specified background, i.e., there the soft modes $\zeta_\uL$ are fixed.
We will use this relation in the following computations.

\subsection{Bispectrum in the squeezed limit}

In order to make this paper self-contained,
we will first review how the bispectrum in the squeezed limit evaluated in the original coordinates is transformed into the one in the rescaled coordinates.
Thus, the results in this subsection are not new.

The bispectrum in the squeezed limit ($k_1, k_2 \gg k_3$) represents a correlation between the power of the hard modes 
and the amplitude of the soft modes.
This can be evaluated by taking the ensemble average of the hard modes under the fixed soft modes ($\approx$ background) first and then
taking the ensemble average of the soft modes~\cite{Creminelli:2004yq, Seery:2005wm}:
\bae{
    \expval{\zeta_{\bfk_1}\zeta_{\bfk_2}\zeta_{\bfk_3}}=\expval{\expval{\zeta_{\bfk_1}\zeta_{\bfk_2}}_\uB\zeta_{\bfk_3}}.}
The rescaled bispectrum is similarly given by
\be
\langle {\bar \zeta}_{\bfk_1} {\bar \zeta}_{\bfk_2} \zeta_{\bfk_3} \rangle = \langle \langle {\bar \zeta}_{\bfk_1} 
{\bar \zeta}_{\bfk_2} \rangle_\uB
\zeta_{\bfk_3} \rangle.
\ee
Relating these equations via Eq.~(\ref{power-b}),  
one obtains a rescaling transformation of the bispectrum as~\cite{Pajer:2013ana}
\be
    {\bar B}_\zeta (k_\uS,k_\uS,k_\uL)
    =B_\zeta (k_\uS,k_\uS,k_\uL)  +(\ns-1) P_\zeta (k_\uS) P_\zeta (k_\uL), \label{bar-B}
\ee
where $k_\uL \ll k_\uS$.
In terms of the $f_{\rm NL}$ parameter defined by Eq.~(\ref{def-fnl}), the above relation recovers Eq.~(\ref{fnl-cons}).
As an important example, for any single field inflation model satisfying the slow-roll conditions, $f_{\rm NL}$ (in the squeezed limit) is
equal to $-5(\ns-1)/12$. 
Thus, ${\bar f_{\rm NL}}$ vanishes in such slow-roll models.

Here, we comment on the local observer effect on the consistency relation of ultra-slow-roll (non-attractor) inflation~\cite{Namjoo:2012aa,Martin:2012pe,Cai:2017bxr}.
In ultra-slow-roll inflation, the final value of $f_{\rm NL}$ depends on the transition 
from an ultra-slow-roll (non-attractor) phase to a standard slow-roll (attractor) one~\cite{Cai:2017bxr, Passaglia:2018ixg}. 
However, because the ultra-slow-roll limit gives rise to the exactly flat spectrum $\ns-1=0$, one obtains $\fNL=\bfNL$ as a general consequence well after the curvature perturbation (and its bispectrum) have converged to a fixed value. Thus if
the final value of $f_{\rm NL}$ coincides with zero, 
the rescaled ${\bar f_{\rm NL}}$ also vanishes as claimed in Ref.~\cite{Bravo:2017gct}. 
But, if a non-zero $f_{\rm NL}$ remains, ${\bar f_{\rm NL}}$ does not necessarily vanish.

\subsection{Trispectrum in the collapsed limit}

The trispectrum in the so-called collapsed limit, 
where all $k_i$ modes are hard (of order $k_\uS$) but $k_{12}, k_{34} \ll k_\uS$ ($k_{ij} =|{\bfk_i}+{\bfk_j}|$),
represents the long-distance correlation of the local variance of the curvature perturbations. 
Like the case of the bispectrum in the squeezed limit, this can be evaluated by taking the ensemble average
of the hard modes under the fixed soft modes ($\approx$ background) first and then
taking the ensemble average of the soft modes~\cite{Seery:2008ax}.
In terms of the curvature perturbation in the rescaled coordinates, we have~\cite{Seery:2008ax} 
\be
\langle {\bar \zeta}_{\bfk_1} {\bar \zeta}_{\bfk_2} {\bar \zeta}_{\bfk_3} {\bar \zeta}_{\bfk_4} \rangle
=\langle \langle {\bar \zeta}_{\bfk_1} {\bar \zeta}_{\bfk_2} \rangle_\uB \langle {\bar \zeta}_{\bfk_3} {\bar \zeta}_{\bfk_4} \rangle_\uB \rangle +\cdots,
\ee
where $\cdots$ denotes the disconnected part, which is irrelevant to our purpose.
Plugging Eq.~(\ref{power-b}) into this equation,
like Eq.~(\ref{bar-B}),
we obtain the trispectrum in the rescaled coordinates in terms of the trispectrum in the original coordinate plus corrections
given by the power spectrum and the bispectrum as
\begin{align}
    {\bar T}_\zeta ({\bfk_1}, {\bfk_2}, {\bfk_3}, {\bfk_4})&=
    T_\zeta ({\bfk_1}, {\bfk_2}, {\bfk_3}, {\bfk_4})+
    (\ns-1) P_\zeta (k_1) B_\zeta (k_3, k_4, k_{34})+(\ns-1) P_\zeta (k_3) B_\zeta (k_1, k_2, k_{12}) \nonumber \\
    &\qquad+{(\ns-1)}^2 P_\zeta (k_1) P_\zeta (k_3) P_\zeta (k_{12}).
\end{align}
In terms of the $\tau_{\rm NL}$ parameter defined by Eq.~(\ref{def-tnl}), the above relation becomes
\be
{\bar \tau_{\rm NL}}=\tau_{\rm NL}+\frac{6}{5} (\ns-1) f_{\rm NL}+\frac{1}{4} {(\ns-1)}^2. \label{transformation-tau}
\ee
Then, using the relation in the original coordinate $\tau_{\rm NL} \ge 36/25 f_{\rm NL}^2$ and Eq.~(\ref{fnl-cons}),
we obtain
\be
{\bar \tau_{\rm NL}} \ge  \frac{36}{25} f_{\rm NL}^2+\frac{6}{5} (\ns-1) f_{\rm NL}+\frac{1}{4} {(\ns-1)}^2=\frac{36}{25} {\bar f_{\rm NL}}^2. 
\ee 
This demonstrates that the original form of the $\fNL$-$\tNL$
relation is not modified by the rescaling of the spatial coordinates.
In particular, just as in the case for the original coordinates, the relation after the rescaling still remains at equality when only a single DoF contributes to the curvature perturbations.
Intuitively, one may attribute the essence of this result to the physical meaning of the $\fNL$-$\tNL$ relation stated below Eq.~(\ref{4-cone}).

The transformation rules (\ref{fnl-cons}) and (\ref{transformation-tau}) also enable us to investigate how other combinations of the NL parameters change under the rescaling of the spatial coordinates.
One can verify 
\be
\btNL-\frac{36}{25}\bfNL^2=\tNL-\frac{36}{25}\fNL^2. \label{sub}
\ee
Thus, the combination $\tNL-\frac{36}{25}\fNL^2$ may be considered as a natural quantity indicating the DoF of the light fields. 
On the other hand, the ratio $A_{\rm NL} \equiv \frac{25}{36} \frac{\tNL}{\fNL^2}$ has been considered as a useful discriminator of the DoF of the light fields in the literature \cite{Suyama:2010uj, Smidt:2010ra}. 
One can verify
\be
{\bar A_{\rm NL}}-1=\frac{A_{\rm NL}-1}{{(1+p)}^2},~~~~~~~~
p=\frac{5}{12}\frac{\ns-1}{\fNL}.
\ee
In principle, $p$ can take any value in $(-\infty, \infty)$ and hence the ratio $A_{\rm NL}$ is not invariant under the rescaling. 
Therefore, to use it as a DoF discriminator, one must also specify which frame is used, particularly
for $\fNL ={\cal O} (\ns-1)$ where the rescaling effect becomes ${\cal O}(1)$.

\section{Summary}

The soft modes of the curvature perturbations are not an observable for a local observer and merely act as to rescale the background locally.
It is known that after the rescaling the non-linearity parameter $\fNL$ characterizing the amplitude of the bispectrum in the squeezed limit is shifted by a correction determined by the power spectrum and the spectral index.
We found that the trispectrum in the collapsed limit,
after the rescaling, is shifted by a correction determined by the combination of the power spectrum, its spectral index, and the bispectrum, by which we were able to express it in terms of another non-linearity parameter $\tNL$ representing the trispectrum in the same limit.
By using this result, we also found that the form of the inequality between $\fNL$ and $\tNL$ is not affected by the rescaling. 
Thus, the role of the inequality as an indicator of the light DoF contributing to the curvature perturbations remains intact.
We have also clarified that the difference between $\tNL$ and $\frac{36}{25} \fNL^2$ is invariant under the rescaling but the ratio of $\tNL$ to $\frac{36}{25} \fNL^2$ is not. 

Before closing this section, let us briefly mention two applications of our results.
Firstly, it is known that non-Gaussianities in the soft limit produce scale-dependent bias of the halo power spectrum~\cite{Dalal:2007cu}.
Given that such a bias represents true correlation between the soft and hard modes, it is the NL parameters in the rescaled coordinates that appear in the bias~\cite{Pajer:2013ana}.
In Ref.~\cite{Gong:2011gx}, it was shown that the combination 
$\tNL-\frac{36}{25}\fNL^2$ produces a particular type of bias, which provides a new observational test of the $\fNL$-$\tNL$ relation.
From Eq.~(\ref{sub}), we can conclude that the result given in Ref.~\cite{Gong:2011gx} 
about the bias from the term $\tNL-\frac{36}{25}\fNL^2$ is not modified.
Secondly, it is known that non-Gaussianities in the soft limit produce large-scale clustering of primordial black holes (PBHs), which directly form from the large-amplitude curvature perturbations upon horizon reentry~\cite{Tada:2015noa, Young:2015kda}.
In Ref.~\cite{Suyama:2019cst}, it was shown that the PBH correlation function at large distance is proportional to $\tNL$. Since PBH formation occurs when the physical size of the rescaled perturbations reenter the Hubble horizon, it is $\btNL$ rather than $\tNL$ that appears in the PBH correlation function.

\section*{Acknowledgments}

We would like to thank Shuichiro Yokoyama for useful comments.
T.~S. was supported by the MEXT Grant-in-Aid for Scientific Research on Innovative Areas No. 17H06359, and No. 19K03864. Y.~T. is supported by JSPS KAKENHI Grants No. JP18J01992 and No. JP19K14707. M.\,Y. is supported in part by JSPS Grant-in-Aid for Scientific Research Numbers 18K18764, Mitsubishi Foundation, and JSPS Bilateral Open Partnership Joint Research Projects.

\bibliographystyle{JHEP}
\bibliography{draft}

\end{document}